\documentclass[letterpaper, 10 pt, conference]{ieeeconf}  

\IEEEoverridecommandlockouts                              

\usepackage[english]{babel}
\usepackage{amsmath, amssymb, amsfonts, bm}
\usepackage{mathrsfs}
\usepackage{textcomp}
\usepackage[mathcal]{euscript}
\usepackage{enumerate}
\usepackage{mathbbol}
\usepackage{color,graphicx,epstopdf}
\usepackage{dsfont}
\usepackage{hyperref}
\usepackage[noadjust]{cite}
\usepackage{caption}
\usepackage{subcaption}
\usepackage{mathtools}
\usepackage{tikz-cd}
\usepackage{mathtools}
\usepackage{algorithm}
\usepackage{algpseudocode}
\usepackage{url}
\usepackage{authblk}
\usepackage{verbatim}
\usepackage{stackengine}
\usepackage{subcaption}

\newtheorem{theorem}{Theorem}
\newtheorem{definition}{Definition}

\newtheorem{remark}{Remark}

\newtheorem{assumption}{Assumption}

\usepackage[left=1.91cm,right=1.91cm,top=1.91cm,bottom=1.91cm]{geometry}
\usepackage{float}

\newcommand{\deltai}{\delta_{i}}
\newcommand{\bdeltai}{\overline{\delta}_{i}}
\newcommand{\tdeltai}{\widetilde{\delta}_{i}}

\newcommand{\bmdelta}{\bm{\delta}}
\newcommand{\ddeltai}{\dot{\delta}_{i}}
\newcommand{\dddeltai}{\ddot{\delta}_{i}}

\newcommand{\bfvm}{\mathbf{|V|}}

\newcommand{\vmi}{|V_{i}|}
\newcommand{\dvmi}{\dot{|V_{i}|}}
\newcommand{\bvmi}{\overline{|V_{i}|}}
\newcommand{\tvmi}{\widetilde{|V_{i}|}}

\newcommand{\vmj}{|V_{j}|}

\newcommand{\tvmj}{\widetilde{|V_{j}|}}

\title{\LARGE \bf
	 Unified Control of Voltage, Frequency and Angle in Electrical Power Systems: A Passivity and Negative-Imaginary based Approach
}

\author{Yijun Chen$^{1}$, Kanghong Shi$^{1}$, Ian R. Petersen$^{1}$, and Elizabeth L. Ratnam$^{1}$
	
	\thanks{$^{1}$The  School of Engineering, The Australian National University, Canberra, Australia, emails: \{yijun.chen, kanghong.shi, ian.petersen, elizabeth.ratnam \}@anu.edu.au.}%
	
	\thanks{This work was supported by the Australian Research Council under grants DP230102443 and LP210200473.}
}

\begin{document}

	\maketitle
	\thispagestyle{empty}
	\pagestyle{empty}

	\begin{abstract}
	This paper proposes a unified methodology for voltage regulation, frequency synchronization, and rotor angle control in power transmission systems considering a one-axis generator model with time-varying voltages.  First, we formulate an output consensus problem with a passivity and negative-imaginary (NI) based control framework. We establish output consensus results for both networked passive systems and networked NI systems. Next, we apply the output consensus problem by controlling large-scale batteries co-located with synchronous generators --- using real-time voltage phasor measurements. By controlling the battery storage systems so as to dispatch real and reactive power, we enable simultaneous control of voltage, frequency, and power angle differences across a transmission network. Validation through numerical simulations on a four-area transmission network confirms the robustness of our unified control framework.
	\end{abstract}

	\section{Introduction}\label{sec:introduction}
    The electric grid that we have known for more than a century must {transform} to meet the global challenge of climate change. The global transformation of power systems is underpinned by the need to displace fossil fuel-fired generation with renewable energy technologies e.g., solar and wind technology {backed by batteries}. This transformation of power systems is in the face of the requirement to continue operating {safely, reliably} and {stably} and achieving this in an {economic way}.
    
    Electric grids of the future need efficient and robust control to regulate voltages, 
    synchronize the grid frequency, and stabilize power angles. Alternative solutions to the conservative engineering approach of building more electricity grid infrastructure (e.g., more power lines) are needed. This conservatism is a direct consequence of a somewhat limited ability to observe and control the grid. The alternative control approaches must support an accelerated pace for the global transformation, as net-zero electricity enables the rapid decarbonization of many sectors, including transport, industry, and buildings.
    
    In recent years, there has been a significant advancement in battery storage systems and associated power electronics.     However, the grid integration of large-scale batteries requires a new framework to understand the interaction between fast-switching power electronics and the dynamic behavior of power transmission networks. Today, at the transmission level, the dynamics of voltage magnitude respond much faster than the rotor angle dynamics. Accordingly, the literature typically separates voltage control from frequency and angle control at the transmission level \cite{sauer2017power,machowski2020power}. However, our paper considers power system models that involve coupling between generator voltage, frequency and rotor angle \cite{bergen1986lyapunov, trip2016internal}.  In contrast to traditional approaches, we propose to use robust feedback control involving large-scale batteries as actuators to decouple the voltage dynamics and the angle dynamics, addressing the simultaneous control of voltage, frequency, and angle.


    Passivity systems theory \cite{desoer2009feedback,khalil2002control} is an important framework in the robust and nonlinear control literature. The work of \cite{hill1977stability} provides stability results for the single-loop negative interconnection of two passive systems. The work of \cite{chopra2012output} establishes output consensus results for networked passive systems, but only employing static controllers. Compared with these prior works, we extend theoretical results for networked systems to encompass general dynamic controllers. 
    
    Negative-imaginary (NI) systems theory \cite{petersen2010feedback,lanzon2008stability,shi2023output} was developed as an approach to the robust control of highly resonant systems. The angle dynamics modeled by swing equations can be regarded as a highly resonant NI system, rendering the application of NI systems theory a suitable approach for frequency and angle control. Our previous work has leveraged networked NI systems theory to establish control frameworks for frequency and angle regulation in electrical power systems \cite{chen2023design,chen2023nonlinear}. However, these efforts have been constrained by the assumption of fixed voltage magnitudes. In this paper, we address the coupling between angle dynamics and voltage dynamics by proposing a unified control framework rooted in both passivity systems theory and NI systems theory.

    In this paper, we address the control of voltage, frequency, and angle for power grids in a unified manner. We consider a one-axis generator model with time-varying voltages. We first formulate an output consensus problem and propose a passivity and NI based feedback approach to control them at once. We establish output consensus results for both
    networked passive systems and networked NI systems. We next apply the proposed feedback approach by using large-scale batteries co-located at synchronous generators as control actuators and using real-time voltage phasor measurements. By controlling the large-scale batteries to dispatch real and reactive power, our approach has three advantages: 1) real and reactive power provided by the large-scale batteries is controlled in a voltage phasor feedback way, enabling the decoupling of the angle dynamics and voltage dynamics; 2) owing to the decoupling, real power based controllers facilitate the synchronization of bus frequencies and the preservation of desired angle differences, while reactive power based controllers assist in regulating bus voltage magnitudes; 3) the proposed control operates in a fully distributed manner, utilizing only local measurement and local communication of voltage angles and voltage magnitudes.

     This paper is organized as follows. Section \ref{sec:preliminary} provides preliminary knowledge on passivity theory and NI systems theory. Section \ref{sec:output_consensus} establishes output consensus results for networked systems. Section \ref{sec:power_systems} presents an application to power transmission systems. Section \ref{sec:simulation} gives simulation results. Section \ref{sec:conclusion} concludes the paper.
 
	\section{Preliminaries}~\label{sec:preliminary}
	Consider a multiple-input multiple-output (MIMO) nonlinear system with the following state-space model:
	\begin{subequations}
		\label{sys:mimo_sp}
		\begin{align}
			\dot{x} &= f(x,u), \label{eq:mimo_state}\\
			y & = h(x) + g(u), \label{eq:mimo_output}
		\end{align}
	\end{subequations}
	where 
	$x \in \mathbb{R}^{n}$ is the state, $u \in \mathbb{R}^{m}$ is the input, $y \in \mathbb{R}^{m}$ is the output, $f:\mathbb{R}^{n} \times \mathbb{R}^{m} \to \mathbb{R}^{n}$ is a Lipschitz continuous function, and $h:\mathbb{R}^{n} \to \mathbb{R}^{m}$ is a class C$^{1}$ function. The admissible inputs are taken to be piecewise continuous and locally square integrable. We impose Assumption~\ref{apt:g} on the input function $g(u)$ and Assumption~\ref{apt:equilibrium} on the system equilibrium.
	
	\begin{assumption}\label{apt:g} 
		The input function $g(u)$ is independent in each input channel, such that
		\begin{equation}\label{eq:independent_input}
			g(u) = [g^{1}(u^{1}), \dots, g^{m}(u^{m})]^{\top},
		\end{equation}
		where each $g^{k}(u^{k})$ is a class $C^{1}$ function with the superscript $k \in \{1,2,\dots,m\}$ representing the $k$th element of the input $u$. Moreover, $g(0) = 0.$
	\end{assumption}
	
	\begin{assumption}\label{apt:equilibrium}
		Without loss of generality, assume $(x^{\ast},u^{\ast}) = (0,0)$ is an equilibrium point of the system~\eqref{sys:mimo_sp}; i.e., $f(0,0)\equiv  0$. Moreover, assume the output at the equilibrium $(0,0)$ is $y^{\ast} \equiv h(0) + g(0) \equiv 0$.
	\end{assumption}
	\medskip
	
	In this paper, we consider systems of the form~\eqref{sys:mimo_sp} which satisfy assumptions that are nonlinear extensions of conventional properties applicable to linear systems as outlined in \cite{shi2023output}. Assumption~\ref{apt:state_output_consistency} is an observability criterion, while Assumption~\ref{apt:input_state_consistency} necessitates that all system inputs exert an influence on the system dynamics.
	
	\begin{assumption}\label{apt:state_output_consistency}
		For any time interval $[t_a,t_b]$ where $t_b>t_a$, the function $h(x)$ remains constant if and only if the state $x$ remains constant. That is $ \dot{h}(x) \equiv 0$ if and only if  $x \equiv \overline x$. Moreover, $h(x) \equiv 0 $ if and only if  $ x\equiv 0$.
	\end{assumption}
	
	\begin{assumption}\label{apt:input_state_consistency}
		For any time interval $[t_a,t_b]$ where $t_b>t_a$, if the state $x$ remains constant, then the input $u$ must also remain constant. That is $x\equiv\overline x$ implies $u\equiv\overline u $. Moreover, $x\equiv 0$ implies $ u\equiv 0$.
	\end{assumption}
	
	\subsection{Passive Systems}
	We review the passivity property and the output strict passivity property for nonlinear MIMO systems \cite{brogliato2007dissipative}.
	\begin{definition}
		\label{def:mimo_passive}
		The system~\eqref{sys:mimo_sp} is said to be 
		passive if there exists a positive semidefinite storage function $S:\mathbb{R}^{n} \to \mathbb{R}$ of class C$^{1}$ such that for any locally integrable input $u$ and solution $x$ to (\ref{eq:mimo_state}), then
		$
			\dot{S}(x)  \leq u^{\top}y,
		$
		for all $t \geq 0$.
	\end{definition}
	
	\begin{definition}
		\label{def:mimo_ospassive}
		The system~\eqref{sys:mimo_sp} is said to be output strictly passive if there exists a positive semidefinite storage function $S:\mathbb{R}^{n} \to \mathbb{R}$ of class C$^{1}$ and a scalar $\epsilon > 0$ such that for any locally integrable input $u$ and solution $x$ to (\ref{eq:mimo_state}), then $
			\dot{S}(x) \leq u^{\top}y - \epsilon\|h(x)\|^{2},
		$
		for all $t \geq 0$.
	\end{definition}
	
	\subsection{Negative-Imaginary Systems}
	We introduce the negative imaginary (NI) property, and output strictly negative imaginary (OSNI) property for nonlinear MIMO systems \cite{shi2023output,chen2023nonlinear}. 
	\begin{definition}
		\label{def:mimo_ni}
		The system~\eqref{sys:mimo_sp} is said to be 
		NI if there exists a positive semidefinite storage function $S:\mathbb{R}^{n} \to \mathbb{R}$ of class C$^{1}$ such that for any locally integrable input $u$ and solution $x$ to (\ref{eq:mimo_state}), then
		 $
			\dot{S}(x) \leq u^{\top}\dot{y},
		$
		for all $t \geq 0$.
	\end{definition}
	
	\begin{definition}
		\label{def:mimo_osni}
		The system~\eqref{sys:mimo_sp} is said to be 
		OSNI if there exists a positive semidefinite storage function $S:\mathbb{R}^{n} \to \mathbb{R}$ of class C$^{1}$ and a scalar $\epsilon > 0$ such that for any locally integrable input $u$ and solution $x$ to (\ref{eq:mimo_state}), then $
			\dot{S}(x) \leq u^{\top}\dot{y} - \epsilon\|\dot{h}(x)\|^{2},
		$
		for all $t \geq 0$.
	\end{definition}
	
	\section{Output Consensus of Networked Systems}\label{sec:output_consensus}
	This section considers a network setting, and output consensus results are presented for the negative feedback interconnection of two networked passive systems and for the positive feedback interconnection of two networked NI systems.
	
	\subsection{Settings for Networked Systems}
	{\bf \noindent Network Setting.} In what follows, we consider a connected and undirected network $\mathcal{G} = (\mathcal{V}, \mathcal{E})$, where $\mathcal{V} = \{1, 2, \dots, N\}$ describes the set of $N$ nodes, and $\mathcal{E} = \{e_{1}, e_{2}, \dots, e_{L}\} \subseteq \mathcal{V} \times \mathcal{V}$ represents the set of $L$ edges connecting the nodes. The index set for edges is denoted by $\mathcal{L}=\{1,2,\dots,L\}.$ Each node is associated with an independent nonlinear plant, while each edge is associated with a nonlinear controller. Each edge takes the outputs of two end nodes as its input, and each node takes the outputs of its connected edges as its input.
	
	Nodes $i$ and $j$ are considered neighboring if there exists an edge $(i, j) \in \mathcal{E}$ connecting them. The set of neighbors for node $i$ is denoted as $\mathcal{N}_{i}$. The structure of the network is represented by the incidence matrix $\mathbf{Q} \in \mathbb{R}^{N \times L}$, where $\mathbf{Q}_{ie}$ is defined as follows:
	\begin{equation*}
		\mathbf{Q}_{ie} =
		\begin{cases}
			1, & \text{if node } i \text{ is the initial node of edge } e,\\
			-1, & \text{if node } i \text{ is the terminal node of edge } e,\\
			0, & \text{if node } i \text{ is not connected to edge } e.
		\end{cases}
	\end{equation*}
	It is noted that a fixed representation of edges is chosen, where each $(i, j)$ or $(j, i)$ can only be chosen once, and ``initial/terminal node'' does not refer to a specific orientation.
	\medskip
	
	{\bf \noindent  Node Plants.} Each node $i \in \mathcal{V}$ is associated with an independent  nonlinear plant $H_{pi}$ described by:
	\begin{subequations}
		\label{sys:networked_h1_sp}
		\begin{align}
			H_{pi} : \quad	\dot{x}_{pi} &= f_{pi}(x_{pi},u_{pi}), \label{eq:networked_h1_state}\\
			y_{pi} & = h_{pi}(x_{pi}), \label{eq:networked_h1_output}
		\end{align}
	\end{subequations}
	where $x_{pi} \in \mathbb{R}^{n_{pi}}$ is the state, $u_{pi} \in \mathbb{R}^{m}$ is the input, $y_{pi} \in \mathbb{R}^{m}$ is the output, $f_{pi}:\mathbb{R}^{n_{pi}} \times \mathbb{R}^{m} \to \mathbb{R}^{n_{pi}}$ is a Lipschitz continuous function, and $h_{pi}:\mathbb{R}^{n_{pi}} \to \mathbb{R}^{m}$ is a class C$^{1}$ function. The admissible inputs are taken to be piecewise continuous and locally square integrable. For a compact expression, we collect the states, inputs and outputs of all nodes --- as represented by the aggregated state vector $X_{p}=[x_{p1}^{\top}, \dots, x_{pN}^{\top}]^{\top} \in \mathbb{R}^{n_{p}}$ with $n_{p} = \sum_{i = 1}^{N}n_{pi}$, the aggregated input vector $U_{p}=[u_{p1}^{\top}, \dots, u_{pN}^{\top}]^{\top} \in \mathbb{R}^{mN}$, and the aggregated output vector $Y_{p}=[y_{p1}^{\top}, \dots, y_{pN}^{\top}]^{\top}\in \mathbb{R}^{mN}$. We denote the aggregated node plants by $\mathcal{H}_{p}$, which is described by 
	\begin{align*}
		\mathcal{H}_{p}: \dot{X}_{p} = \begin{bmatrix}
			f_{p1}(x_{p1},u_{p1})\\
			\vdots\\
			f_{pN}(x_{pN},u_{pN})
		\end{bmatrix},
		Y_{p} = \begin{bmatrix}
			h_{p1}(x_{p1})\\
			\vdots\\
			h_{pN}(x_{pN})
		\end{bmatrix}.
	\end{align*}
	We further denoted the storage function for each node plant $H_{pi}, i \in \mathcal{V}$ by $S_{pi}$. The storage functions for the aggregated node plants $\mathcal{H}_{p}$ is chosen as $S_{p} = \sum_{i \in \mathcal{V}}S_{pi}$.
	\medskip
	
	{\bf \noindent Edge Controllers.} Each edge $e_{l} \in \mathcal{E}$ with $l \in \mathcal{L}$ is deployed with a  nonlinear controller described by 
	\begin{subequations}
		\label{sys:networked_h2_sp}
		\begin{align}
			H_{cl} : \quad	\dot{x}_{cl} &= f_{cl}(x_{cl},u_{cl}), \label{eq:networked_h2_state}\\
			y_{cl} & = h_{cl}(x_{cl}) + g_{cl}(u_{cl}), \label{eq:networked_h2_output}
		\end{align}
	\end{subequations}
	where $x_{cl} \in \mathbb{R}^{n_{cl}}$ is the state, $u_{cl} \in \mathbb{R}^{m}$ is the input, $y_{cl} \in \mathbb{R}^{m}$ is the output, $f_{cl}:\mathbb{R}^{n_{cl}} \times \mathbb{R}^{m} \to \mathbb{R}^{n_{cl}}$ is a Lipschitz continuous function, and $h_{cl}:\mathbb{R}^{n_{cl}}  \to \mathbb{R}^{m}$ is a class C$^{1}$ function. Assumption~\ref{apt:g} is assumed for the input functions $g_{cl}(u_{cl}), l \in \mathcal{L}$. The admissible inputs are taken to be piecewise continuous and locally square integrable.  For a compact expression, we collect the states, the inputs and the outputs of all edges into the aggregated state vector $X_{c}=[x_{c1}^{\top}, \dots, x_{cL}^{\top}]^{\top} \in \mathbb{R}^{n_{c}}$ with $n_{c} = \sum_{l \in \mathcal{L}}n_{cl}$, the aggregated input vector $U_{c}=[u_{c1}^{\top}, \dots, u_{cL}^{\top}]^{\top} \in \mathbb{R}^{mL}$, and the aggregated output vector 
	$
		Y_{c}=[y_{c1}^{\top}, \dots, y_{cL}^{\top}]^{\top} = \Pi_{cx}(X_{c}) + \Pi_{cu}(U_{c}) \in \mathbb{R}^{mL}, 
	$
	where 
	$
			\Pi_{cx}(X_{c}) = [h_{c1}(x_{c1})^{\top}, \dots, h_{cL}(x_{cL})^{\top}]^{\top} \in \mathbb{R}^{mL},$ and $
			\Pi_{cu}(U_{c}) = [g_{c1}(u_{c1})^{\top}, \dots, g_{cL}(u_{cL})^{\top}]^{\top} \in \mathbb{R}^{mL}.
		$
	We denote the aggregated nonlinear controllers by $\mathcal{H}_{c}$, which are described by 
	\begin{align*}
		\mathcal{H}_{c}: \dot{X}_{c} = \begin{bmatrix}
			f_{c1}(x_{c1},u_{c1})\\
			\vdots\\
			f_{cL}(x_{cL},u_{cL})
		\end{bmatrix},
		Y_{c} = \begin{bmatrix}
			h_{c1}(x_{c1}) + g_{c1}(u_{c1})\\
			\vdots\\ 
			h_{cL}(x_{cL}) + g_{cL}(u_{cL})
		\end{bmatrix}.
	\end{align*}
	We further denoted the storage function for each node plant $H_{cl}, l \in \mathcal{L}$ by $S_{cl}$. The storage function for the aggregated edge controller $\mathcal{H}_{c}$ is chose as $S_{c} = \sum_{l \in \mathcal{L}}S_{cl}$.
	\medskip
	
	{\bf \noindent Output Feedback Control Framework.} The objective of our control problem is to achieve output consensus for each node in the network. We now define local output consensus. 
	
	\begin{definition}[Output Consensus]\label{def:output_consensus}
		A distributed output feedback control law achieves local output feedback consensus for a networked system if there exists an open domain $\mathcal{D}_{c} \subset \mathbb{R}^{n_{p}\times n_{c}}$ containing the origin such that $\lim_{t \to \infty}\|y_{pi}(t) - y_{pj}(t)\| = 0,$ for all $ i,j \in \mathcal{V},$ for all initial conditions $(X_{p}(0),X_{c}(0)) \in \mathcal{D}_{c}.$ 
	\end{definition}
	
	\medskip

\begin{figure}[bt]
		\centering
		\includegraphics[width=\linewidth]{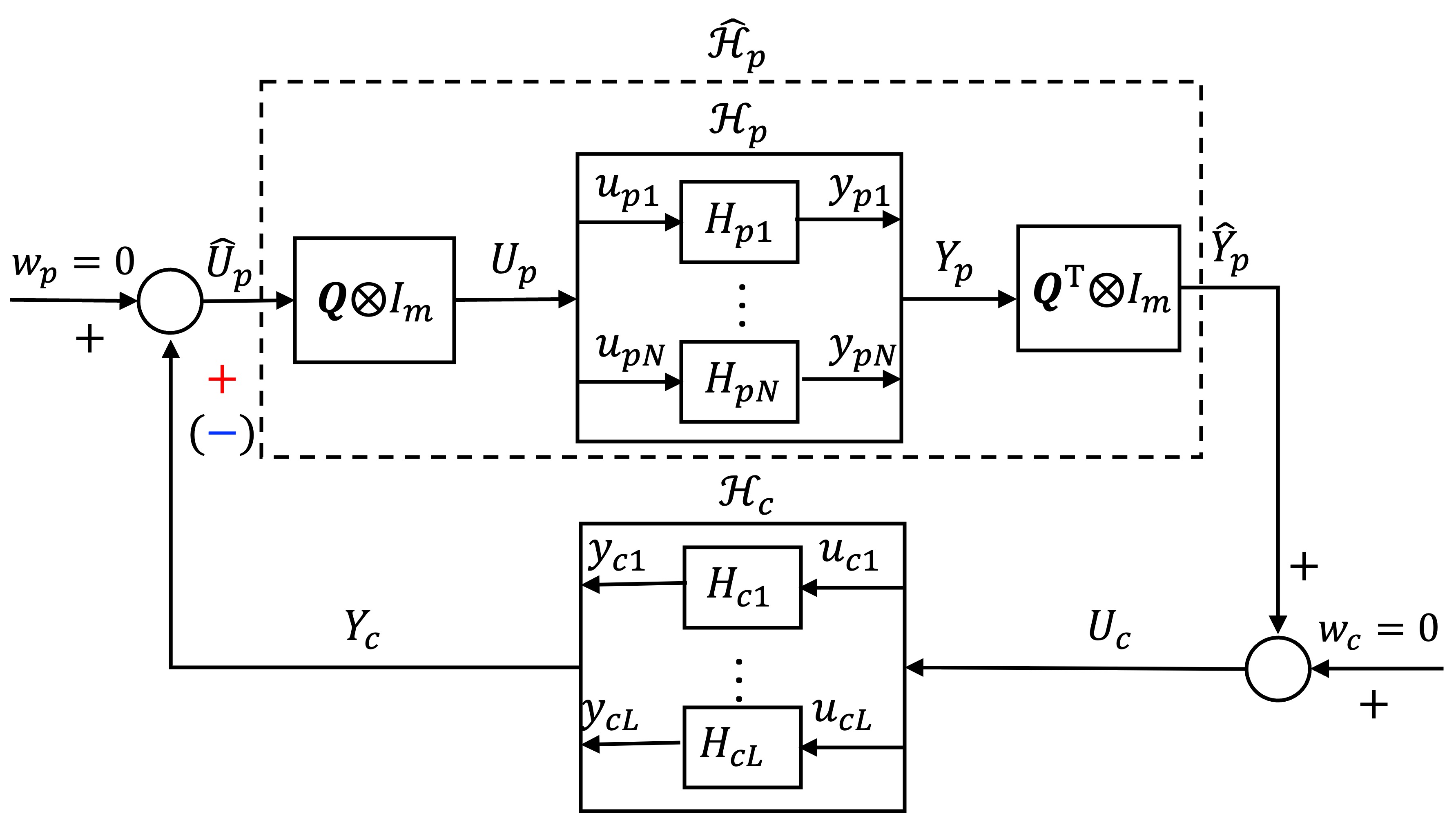}
		\caption{The positive (negative) feedback interconnection of nonlinear plants $\mathcal{H}_{p}$ and nonlinear edge controllers $\mathcal{H}_{c}$ based on the underlying network, where the feedback sign is `$+$' in red (`$-$' in blue).}
		\label{fig:fb}
     \vspace{-0.5em}
	\end{figure}

	As depicted in Fig.~\ref{fig:fb},  two distributed output feedback control frameworks naturally arise based on the underlying network: (1) negative feedback interconnection with feedback sign `$-$' in blue; (2) positive feedback interconnection with feedback sign `$+$' in red. We denote the networked node plants by $\widehat{\mathcal{H}}_{p} = (\mathbf{Q}^{\top} \otimes I_{m})\mathcal{H}_{p}(\mathbf{Q}\otimes I_{m})$, whose input and output are denoted by $\widehat{U}_{p}$ and $\widehat{Y}_{p}$, respectively.  For the networked node plants $\widehat{\mathcal{H}}_{p}$, the storage function is chosen as the same for the aggregated node plants $\mathcal{H}_{p}$; i.e., $\widehat{S}_{p} = S_{p} = \sum_{i \in \mathcal{V}}S_{pi}.$  The relation between $U_{p}$ and $\widehat{U}_{p}$, as well as between $Y_{p}$ and $\widehat{Y}_{p}$, is expressed by
	\begin{subequations}\label{eq:post-processed_input_output}
		\begin{align}
			& U_{p} = (\mathbf{Q} \otimes I_{m})\widehat{U}_{p}, \label{eq:U_p_U_hat_p}
			\\
			& \widehat{Y}_{p} \equiv (\mathbf{Q}^{\top}\otimes I_{m}) Y_{p}.  \label{eq:Y_hat_p_Y_p}
		\end{align}
	\end{subequations}
	We denote the negative feedback interconnection by $(\widehat{\mathcal{H}}_{p}, \mathcal{H}_{c})^{-}$.  The relation between the inputs and the outputs of the negative feedback system $(\widehat{\mathcal{H}}_{p}, \mathcal{H}_{c})^{-}$ is described by 
	\begin{align}\label{eq:negative_input_output}
		\widehat{U}_{p} \equiv -Y_{c} \  \text{ and } \ 
		\widehat{Y}_{p} \equiv U_{c}.
	\end{align}
	We denote the positive feedback interconnection by $(\widehat{\mathcal{H}}_{p}, \mathcal{H}_{c})^{+}$.  The relation between the inputs and the outputs of the positive feedback system $(\widehat{\mathcal{H}}_{p}, \mathcal{H}_{c})^{+}$ is described by 
	\begin{align}\label{eq:positive_input_output}
		\widehat{U}_{p} \equiv Y_{c} \  \text{ and } \ 
		\widehat{Y}_{p} \equiv U_{c}.
	\end{align}
	
	Under both frameworks, the overall systems operate in a distributed manner. Each edge controller $l \in \mathcal{L}$ takes the difference between the outputs of the neighbouring nodes $i$ and $j$ as its input,
	$u_{cl} = \sum_{k = 1}^{N}q_{kl} y_{pk} = y_{pi} - y_{pj},$
	where $q_{kl}$ represents the $k$th element in the $l$th column of the incidence matrix $\mathbf{Q}$, and the node $i$ and the node $j$ are the initial node and the terminal node of the edge $e_{l}$, respectively. 
	In the case of a positive (negative) feedback interconnection, each node plant $i \in \mathcal{V}$ takes the sum of the  outputs from all its connected edge controllers as its input,
	$u_{pi} =  \sum_{l = 1}^{ L} q_{il}y_{cl}$ ($u_{pi} =  -\sum_{l = 1}^{ L} q_{il}y_{cl}),$
	where $q_{il}$ is the $l$th element in the $i$th row of the incidence matrix $\mathbf{Q}.$ 
	
	\subsection{Main Results}
	\subsubsection{\bf Output Consensus for Networked Passive Systems}
	
	Consider passive node plants $H_{pi}, i \in \mathcal{V}$. Also consider output strictly passive edge controllers $H_{cl}, l \in \mathcal{L}$. For the negative feedback system $(\widehat{\mathcal{H}}_{p}, \mathcal{H}_{c})^{-}$, a candidate Lyapunov function is selected as 
	\begin{equation}\label{eq:negative_Lyapunov}
		W^{-} = \sum_{i\in \mathcal{V}}S_{pi}(x_{pi}) + 	\sum_{l\in \mathcal{L}}S_{cl}(x_{cl}).
	\end{equation}
	\begin{assumption}\label{apt:negative_networked_local_domain}
		There exists an open domain $\mathcal{D} \subset \mathbb{R}^{n_{p}} \times \mathbb{R}^{n_{c}}$ such that the candidate Lyapunov function~\eqref{eq:negative_Lyapunov} is positive definite.
	\end{assumption}
	
	\medskip
	
	The following theorem establishes an output consensus result for the negative feedback system $(\widehat{\mathcal{H}}_{p},\mathcal{H}_{c})^{-}$.
	
	\begin{theorem}
		\label{thm:negative_feedback}
		Consider passive node plants $H_{pi}, i \in \mathcal{V}$. Also consider output strictly passive edge controllers $H_{cl}, l \in \mathcal{L}$. Suppose Assumptions~\ref{apt:g},~\ref{apt:equilibrium},~\ref{apt:state_output_consistency}, and~\ref{apt:input_state_consistency} hold for all node plants $H_{pi}, i \in \mathcal{V}$ and edge controllers $H_{cl}, l \in \mathcal{L}$.   Further, consider the negative feedback system $(\widehat{\mathcal{H}}_{p}, \mathcal{H}_{c})^{-}$.  Suppose  Assumption~\ref{apt:negative_networked_local_domain} holds for the negative feedback system $(\widehat{\mathcal{H}}_{p}, \mathcal{H}_{c})^{-}$. Then, local output consensus is achieved.
	\end{theorem} 
	\medskip
	
	{\it \noindent Proof.} First, according to Assumption~\ref{apt:negative_networked_local_domain}, the candidate Lyapunov function~\eqref{eq:negative_Lyapunov} of the negative feedback system $(\widehat{\mathcal{H}}_{p}, \mathcal{H}_{c})^{-}$ is positive definite on an open domain $\mathcal D$. Second, we  analyze the time derivative of the candidate Lyapunov function~\eqref{eq:negative_Lyapunov}:
	\begin{subequations}\label{eq:dot_W-}
		\begin{align}
			\dot{W}^{-} 
			\leq&U_{p}^{\top}Y_{p}+U_{c}^{\top}Y_{c}- \sum_{l\in \mathcal{L}}\epsilon_{cl}\| h_{cl}(x_{cl}) \|^{2}\label{eq:c}\\
			=&U_{p}^{\top}Y_{p}-\widehat{Y}_{p}^{\top}\widehat{U}_{p}- \sum_{l\in \mathcal{L}}\epsilon_{cl}\| h_{cl}(x_{cl}) \|^{2} \label{eq:d}\\
			= & U_{p}^{\top}Y_{p} - U_{p}^{\top}Y_{p} - \sum_{l\in \mathcal{L}}\epsilon_{cl}\| h_{cl}(x_{cl}) \|^{2} \label{eq:e}\\
			\leq& -\epsilon_{c\min}\|\Pi_{cx}\|^{2}  \leq 0,
		\end{align}
	\end{subequations}
	where $\epsilon_{c\min} = \min\{\epsilon_{c1}, \dots, \epsilon_{cL}\} > 0.$ The inequality~\eqref{eq:c} comes from the fact that each node plant $H_{pi}, i \in \mathcal{V}$ is passive and each edge controller $H_{cl}, l \in \mathcal{L}$ is output strictly passive. The equality~\eqref{eq:d} follows from the input-output relation~\eqref{eq:negative_input_output}, while the equality~\eqref{eq:e} follows from Eqs.~\eqref{eq:U_p_U_hat_p} and~\eqref{eq:Y_hat_p_Y_p}. Therefore, the negative feedback system $(\widehat{\mathcal{H}}_{p}, \mathcal{H}_{c})$ is at least locally stable in the sense of Lyapunov. 
	
	Equation~\eqref{eq:dot_W-} implies that $\dot{W}^{-}$ can remain zero only if $\Pi_{cx}$ remains zero. According to Assumption~\ref{apt:state_output_consistency}, it holds that $\Pi_{cx} \equiv 0$ implies $ X_{c} \equiv 0$. According to Assumption~\ref{apt:input_state_consistency}, it holds that $ X_{c} \equiv 0$ implies $ U_{c} \equiv 0$. From the input-output relation~\eqref{eq:negative_input_output}, we obtain $\widehat{Y}_{p} \equiv 0$; i.e. local output consensus is achieved. Therefore, we conclude that $\dot{W}^{-}$ cannot remain zero unless local output consensus is reached.  \hfill $\square$

	\begin{remark}
		Theorem~\ref{thm:negative_feedback} extends prior research, such as \cite{hill1977stability}, which concentrated solely on stability in the single-loop scenario, and \cite{chopra2012output}, which addressed output consensus problem for networked passive systems exclusively with static controllers. Theorem~\ref{thm:negative_feedback} establishes stability and output consensus results for networked passive systems with dynamic controllers. Specifically, Theorem~\ref{thm:negative_feedback} is applicable when each controller $\mathcal{H}_{cl},l \in \mathcal{L}$ reduces to a static controller $y_{cl} = g_{cl}(u_{cl})$, which is consistent with \cite{chopra2012output}.
	\end{remark}
	
	\subsubsection{\bf Output Consensus for Networked NI Systems}
	Consider NI node plants $H_{pi}, i \in \mathcal{V}$. Also consider OSNI edge controllers $H_{cl}, l \in \mathcal{L}$. 
	For the positive feedback system $(\widehat{\mathcal{H}}_{p}, \mathcal{H}_{c})^{+}$, a candidate Lyapunov function is selected as 
	\begin{align}
		\label{eq:positive_Lyapunov}
		W^{+} = & \sum_{i\in \mathcal{V}}S_{pi}(x_{pi}) + 	\sum_{l\in \mathcal{L}}S_{cl}(x_{cl}) - \widehat{Y}_{p}^{\top}\Pi_{cx} \notag\\
		& - \sum_{k=1}^{mL}\int_{0}^{\widehat{Y}_{p}^{k}}\Pi_{cu}^{k}(\xi^{k})d\xi^{k}.
	\end{align}
	
	\begin{assumption}\label{apt:positive_networked_local_domain}
		There exists an open domain $\mathcal{D} \subset \mathbb{R}^{n_{p}} \times \mathbb{R}^{n_{c}}$ such that the candidate Lyapunov function~\eqref{eq:positive_Lyapunov}is positive definite.
	\end{assumption}
	
	In light of the stability results in \cite{shi2023output}, we impose comparable assumptions for the system $\widehat{\mathcal{H}}_{p}$ and the system $\mathcal{H}_{c}$. 
	\begin{assumption}\label{apt:networked_positive_uh}
		For the system $\widehat{\mathcal{H}}_{p}$ with a constant input $\overline{\widehat{U}}_{p}$ which results in a constant output $\overline{\widehat{Y}}_{p}$, then $\overline{\widehat{U}}_{p}^{\top}\overline{\widehat{Y}}_{p} \geq 0.$ 
	\end{assumption}
	
	\begin{assumption}\label{apt:negative_uy}
		For a system $\mathcal{H}_{c}$ with a constant input $\overline{U}_{c}$ which results in a constant output $\overline{Y}_{c}$, then $\overline{U}_{c}^{\top}\overline{Y}_{c}\leq -\gamma_{c}\| \overline{U}_{c}\|^{2}$ with $\gamma_{c} >0.$ 
	\end{assumption}
	\medskip
	
	The following theorem establishes an output consensus result for the positive feedback system $(\widehat{\mathcal{H}}_{p},\mathcal{H}_{c})^{+}$.
	
	\begin{theorem}
		\label{thm2:positive_feedback}
		Consider NI node plants $H_{pi}, i \in \mathcal{V}$. Suppose Assumptions~\ref{apt:equilibrium},~\ref{apt:state_output_consistency} and~\ref{apt:input_state_consistency} hold for the NI node plants, and Assumption~\ref{apt:networked_positive_uh} holds for the networked node plants $\widehat{\mathcal{H}}_{p}$. Also consider OSNI edge controllers $H_{cl}, l \in \mathcal{L}$. Suppose Assumptions~\ref{apt:g},~\ref{apt:equilibrium},~\ref{apt:state_output_consistency},~\ref{apt:input_state_consistency}, and \ref{apt:negative_uy} hold for the OSNI edge controllers.   Further, consider the positive feedback system $(\widehat{\mathcal{H}}_{p}, \mathcal{H}_{c})^{+}$. Suppose  Assumption~\ref{apt:positive_networked_local_domain} holds for the positive feedback system $(\widehat{\mathcal{H}}_{p}, \mathcal{H}_{c})^{+}$. Then, local output consensus is achieved.
	\end{theorem} 
	\medskip
	{\it Proof.} First, according to Assumption~\ref{apt:positive_networked_local_domain}, the candidate Lyapunov function~\eqref{eq:positive_Lyapunov} of the positive feedback system $(\widehat{\mathcal{H}}_{p}, \mathcal{H}_{c})^{+}$ is positive definite. Second, we analyze the time derivative of the candidate Lyapunov function~\eqref{eq:positive_Lyapunov}:
	\begin{align}\label{eq:dot_W+}
		\dot{W}^{+}
		\leq & - \epsilon_{c\min}\|\dot{\Pi}_{cx}\|^{2} \leq  0, 
	\end{align}
	where $\epsilon_{c\min} = \min\{\epsilon_{c1}, \dots, \epsilon_{cL}\} > 0.$  Therefore, the positive feedback system $(\widehat{\mathcal{H}}_{p}, \mathcal{H}_{c})^{+}$ is at least locally stable in the sense of Lyapunov. 
	
	Equation~\eqref{eq:dot_W+} implies that $\dot{W}^{+}$ can remain zero only if $\dot{\Pi}_{cx}$ remains zero. For the aggregated edge controllers $\mathcal{H}_{c}$,
	$
	\dot{\Pi}_{cx} \equiv 0$ implies a steady state $\overline{X}_{c}$, constant input $\overline{U}_{c}$, and constant output $\overline{Y}_{c}$ from Assumptions~\ref{apt:state_output_consistency} and~\ref{apt:input_state_consistency}. According to the input-output relation~\eqref{eq:positive_input_output}, constant input  $\overline{U}_{c}$ and constant output $\overline{Y}_{c}$ of the aggregated edge controllers $\mathcal{H}_{c}$ implies constant output $\overline{\widehat{Y}}_{p}$ and constant input $\overline{\widehat{U}}_{p}$ of  the networked node plants $\widehat{\mathcal{H}}_{p}$. 
	We conclude that the networked node plants $\widehat{\mathcal{H}}_{p}$ subject to constant input $\overline{\widehat{U}}_{p}$ has constant output $\overline{\widehat{Y}}_{p}$ and the aggregated edge controllers subject to constant input $\overline{U}_{c}$ has constant output $\overline{Y}_{c}$; i.e., Assumptions~\ref{apt:networked_positive_uh} and~\ref{apt:negative_uy} are satisfied.  We have the following facts that
	$\overline{\widehat{U}}_{p}^{\top}\overline{\widehat{Y}}_{p} \equiv \overline{U}_{c}^{\top}\overline{Y}_{c}$, $
			\overline{\widehat{U}}_{p}^{\top}\overline{\widehat{Y}}_{p} \geq 0$, and $\overline{U}_{c}^{\top}\overline{Y}_{c}\leq -\gamma_{c}\| \overline{U}_{c}\|^{2},$ which   can only be valid when $\overline{U}_{c} \equiv \overline{\widehat{Y}}_{p} \equiv 0$; i.e., local output consensus is achieved. Therefore, we conclude that $\dot{W}^{+}$ cannot remain zero unless local output consensus is achieved. 
	\hfill $\square$
	\medskip
	
	\section{Application to Power Transmission Systems}\label{sec:power_systems}
	In this section, we apply the above theoretical results to the practical problems of frequency synchronization, angle difference preservation, and voltage regulation in electrical power systems.
	
	\subsection{One-axis Generator Model}
	Consider a transmission network, whose topology is represented by a connected and undirected graph $\mathcal{G} = (\mathcal{V},\mathcal{E})$. The transmission network consists of $N$ nodes representing synchronous generator buses and $L$ edges representing transmission lines.  The nominal frequency of the transmission network is denoted by $\omega_{nom}$. Without loss of generality, the nominal voltage magnitude for all buses is defined by $V_{nom}$.   Each synchronous generator bus $i \in \mathcal{V}$ is associated with a voltage phasor $V_{i} = \vmi \angle \deltai,$ where $\deltai$ is the voltage angle and $\vmi$ is the voltage magnitude. The frequency of each synchronous generator bus is denoted by $\omega_{i}.$  The relation between the voltage angle and the bus frequency is described by $\omega_{i} = \dot{\delta}_{i} + \omega_{nom}.$ 
	
	In contrast to traditional power systems, we have incorporated large-scale batteries equipped at each synchronous generator, which can be utilized to provide real and reactive power  to synchronize bus frequencies, maintain angle differences, and  regulate voltage magnitudes. Throughout the paper, the complex power at each synchronous generator bus $i \in \mathcal{V}$ will be denoted by $P_{i} + j Q_{i}$, where $P_{i}$ and $Q_{i}$ refer to real power and reactive power, respectively. Distinct superscripts are used when referring to the corresponding power sources.  
	
	The synchronous generators are characterized by a one-axis generator model; i.e., this model considers swing equations under the effects of field flux decays. The dynamics of each synchronous generator $i \in \mathcal{V}$ are described by \cite{bergen1986lyapunov}:
	\begin{subequations}
		\begin{align}
			&M_{i} \dddeltai + D_{i} \ddeltai = P_{i}^{G} - P_{i}^{E}(\bmdelta,\bfvm) + P_{i}^{ST}, \label{eq:angle_dynamics}\\
			&\frac{T'_{doi}}{X_{di} - X_{di}'} \vmi \dvmi = Q_{i}^{G}(\vmi) - Q_{i}^{E}(\bmdelta,\bfvm) + Q_{i}^{ST}, \label{eq:voltage_dynamics}
		\end{align}
	\end{subequations}
	where $P_{i}^{G}$ is the fixed mechanical power inputs, $Q_{i}^{G}$ can be regarded as the reactive power supplied by the exciter to the generator bus, $P_{i}^{E}$, $Q_{i}^{E}$ are the total real, reactive power flow from the $i$-th generator bus via transmission lines, and $P_{i}^{ST}$, $Q_{i}^{ST}$ are the total real, reactive power output from the large-scale battery. Writing the angle dynamics~\eqref{eq:angle_dynamics} and  voltage dynamics~\eqref{eq:voltage_dynamics}  in this manner provides us with a clear idea that, although angle dynamics and voltage dynamics are coupled, we may use real power to regulate frequency and angle, while using reactive power to regulate voltage magnitude. The expressions for $P_{i}^{E}$, $Q_{i}^{G}$, and $Q_{i}^{E}$  for each synchronous generator bus $i \in\mathcal{V}$ are described by \cite{bergen1986lyapunov}
	\begin{subequations}
		\begin{align}
			&P_{i}^{E}(\bmdelta,\bfvm) = \sum_{j =1}^{N}B_{ij} \vmi \vmj\sin(\delta_{i} - \delta_{j}), \label{eq:PE}\\
			&Q_{i}^{G}(\vmi) = \frac{\vmi (E^{ex}_{i} - \vmi)}{X_{di} - X_{di}'},  \label{eq:QG}\\
			&Q_{i}^{E}(\bmdelta,\bfvm) = - \sum_{j =1}^{N} B_{ij} \vmi \vmj\cos(\delta_{i} - \delta_{j}),  \label{eq:QE}
		\end{align}
	\end{subequations}
	where $\bmdelta = [\delta_{1}, \dots, \delta_{N}]^{\top}$ and $\bfvm = [|V_{1}|, \dots, |V_{N}|]^{\top}$.
	
	In~Eqs.~\eqref{eq:angle_dynamics}-\eqref{eq:QE},  the notation used for each synchronous machine $i \in \mathcal{V}$ is summarized in Table~\ref{eq:notations}. 
	\begin{table}[h]
		\normalsize
		\centering
		\begin{tabular}{ll}
			\hline
			& State variables\\
			\hline
			$\vmi$&Machine internal voltage magnitude\\
			$\deltai$&Machine internal voltage angle\\
            $\omega_{i}$& Machine frequency\\
			\hline
			& System parameters\\
			\hline
			$M_{i}$&Machine inertia constant\\
			$D_{i}$&Machine damping constant\\
			$B_{ij}$&Transmission line $(i,j)$'s susceptance\\
			$T'_{doi}$&Direct axis transient open-circuit time constant\\
			$X_{di}$&Direct axis synchronous reactance \\
			$X_{di}'$&Direct axis transient reactance \\
			\hline
			& Control inputs\\
			\hline
			$P_{i}^{ST}$&Battery storage real power output\\
			$Q_{i}^{ST}$&Battery storage reactive power output\\
			\hline
			& Other fixed inputs\\
			\hline
			$P^{G}_{i}$&Mechanical real power input\\
			$E^{ex}_{i}$&Excitation voltage magnitude\\
			\hline
		\end{tabular}
		\caption{The notation used for each synchronous machine.}
		\label{eq:notations}
        \vspace{-0.5em}
	\end{table}
	
	For each synchronous generator bus $i \in \mathcal{V}$ at equilibrium, the steady-state angle is denoted by $\bdeltai$ and the steady-state voltage magnitude is denoted by $\bvmi = V_{nom}$.  We define the angle deviation and voltage magnitude deviation from the equilibrium by $\tdeltai = \deltai - \bdeltai$ and $\tvmi = \vmi - \bvmi$, respectively. We also define $\delta_{ij} = \delta_{i} - \delta_{j}$, $\overline{\delta}_{ij} = \overline{\delta}_{i} - \overline{\delta}_{j}$, and $\widetilde{\delta}_{ij} = \widetilde{\delta}_{i} - \widetilde{\delta}_{j}$.
	
	Suppose that large-scale batteries are not supplying apparent power to the electric grid at the equilibrium, i.e., $\overline{P}^{ST}_{i} = 0$ and $\overline{Q}^{ST}_{i} = 0$. Taking into account the condition at the equilibrium, the angle dynamics~\eqref{eq:angle_dynamics} and the voltage dynamics~\eqref{eq:voltage_dynamics} can be rewritten in terms of angle deviations and voltage magnitude deviations: 
	\begin{subequations}
		\begin{align}
			&\ddot{\widetilde{\delta}}_{i} = -\frac{D_{i}}{M_{i}}\dot{\widetilde{\delta}}_{i} + \frac{1}{M_{i}}u_{pi}^{\delta}(\bmdelta, \bfvm, P_{i}^{ST}), \label{eq:decouple_angle_dynamics}\\
			&\dot{\tvmi} = -\frac{\alpha_{i}}{\gamma_{i}}\tvmi + \frac{1}{\gamma_{i}}u_{pi}^{V}(\bmdelta, \bfvm, Q_{i}^{ST}), \label{eq:decouple_voltage_dynamics}
		\end{align}
	\end{subequations}
	where $\alpha_{i} = \frac{1}{X_{di} - X_{di}'} - B_{ii}$ and $\gamma_{i} = \frac{T'_{doi}}{X_{di} - X_{di}'}$. The inputs $u_{i}^{\delta}$ and $u_{i}^{V}$ are expressed by
		\begin{align*}
			u_{pi}^{\delta} &= P_{i}^{ST} + \sum_{j \in \mathcal{N}_{i}}B_{ij}\big(V_{nom}^{2}\sin\overline{\delta}_{ij} 
			- \vmi \vmj \sin \delta_{ij} \big),\\
			u_{pi}^{V}  &= \frac{Q_{i}^{ST}}{\vmi} - \sum_{j \in \mathcal{N}_{i}}B_{ij}\big( V_{nom}\cos\overline{\delta}_{ij} 
			- \vmj \cos\delta_{ij}\big).
		\end{align*}
	
	\begin{remark}
		In transmission systems, reactance is typically much greater than resistance, and steady-state reactance is typically much higher than transient reactance, i.e., $X_{di} - X_{di'} > 0.$ Also, the self-susceptance is usually less than zero, i.e., $B_{ii} <0$ \cite{trip2016internal}. Accordingly, $\frac{\alpha_{i}}{\gamma_{i}} > 0.$
	\end{remark}
	
\begin{remark}
In this paper, we assume that all generator buses are equipped with large-scale batteries. In other words, once we design $u_{pi}^{\delta}$ and $u_{pi}^{V}$, the control actions exerted by the large-scale batteries can be described by:
	\begin{align*}
		P_{i}^{ST} &= u_{pi}^{\delta} - \sum_{j \in \mathcal{N}{i}}B_{ij}\left(V_{nom}^{2}\sin\overline{\delta}_{ij}
		- \vmi \vmj \sin \delta_{ij} \right),\\
		Q_{i}^{ST} &= {\vmi} \big( u_{pi}^{V} + \sum_{j \in \mathcal{N}{i}}B_{ij}\left( V_{nom}\cos\overline{\delta}_{ij}
		- \vmj \cos\delta_{ij} \right) \big).
	\end{align*}
This enables us to enhance power system transient stability and robustness by utilizing the real and reactive power outputs of the large-scale batteries.
\end{remark}

	
	\subsection{Feedback Decoupling and Linearization}
	We assume real-time measurements of voltage phasor for each generator bus are available and that all generator buses have large-scale batteries that can provide real and reactive power as control actuators. We investigate the application of voltage phasor feedback control in which the angle dynamics and the voltage dynamics can be decoupled, and utilize the results established in Section~\ref{sec:output_consensus} to guarantee the synchronization of bus frequencies, the preservation of angle differences, and the regulation of voltage magnitudes.
	
\begin{figure}[bt]
		\centering
		\includegraphics[width=0.88\linewidth]{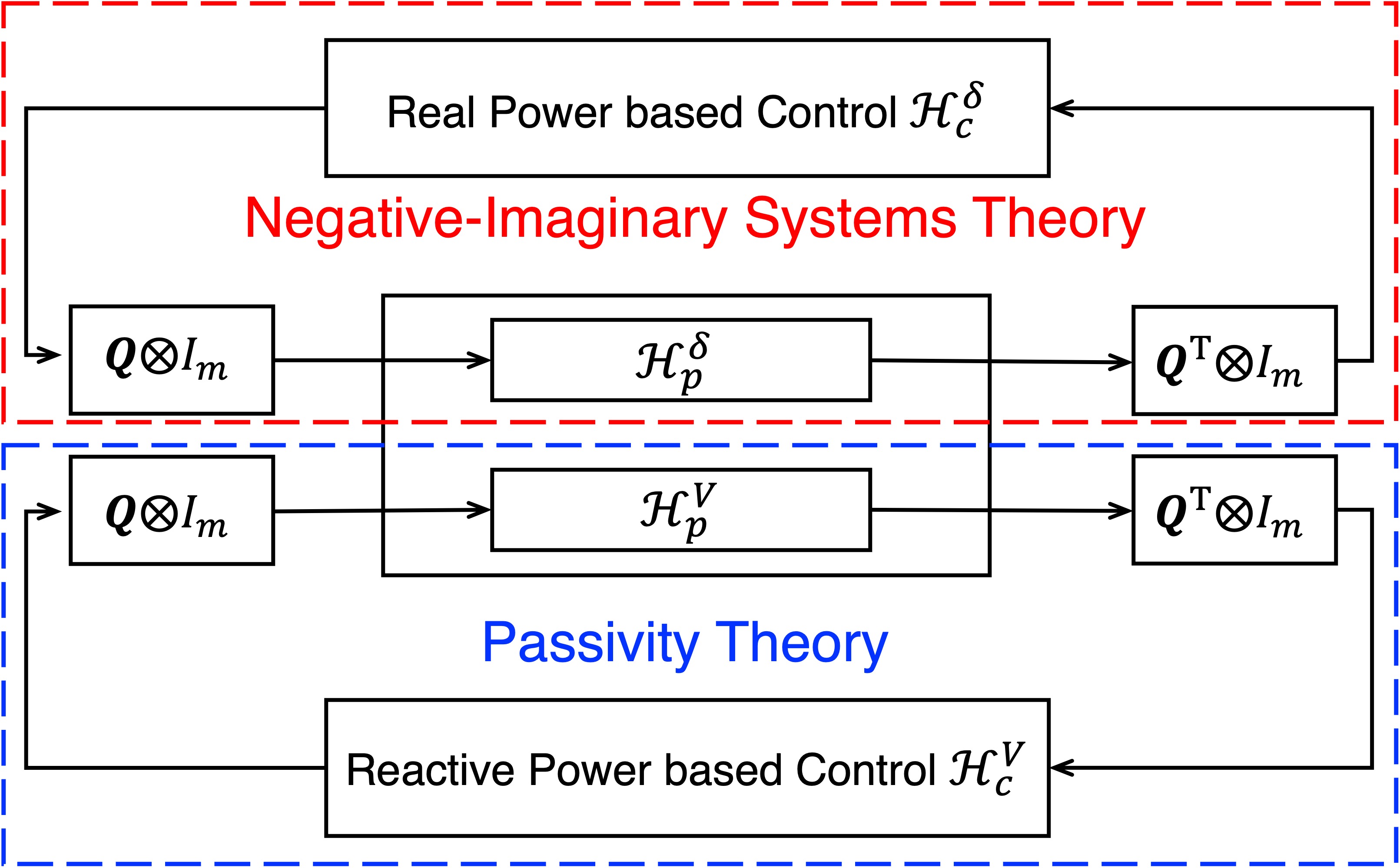}
		\caption{The feedback control framework, where the angle dynamics and the voltage dynamics can be decoupled.}
		\label{fig:overall_fb}
	\end{figure}

\begin{figure}[bt]
		\centering
		\includegraphics[width=0.95\linewidth]{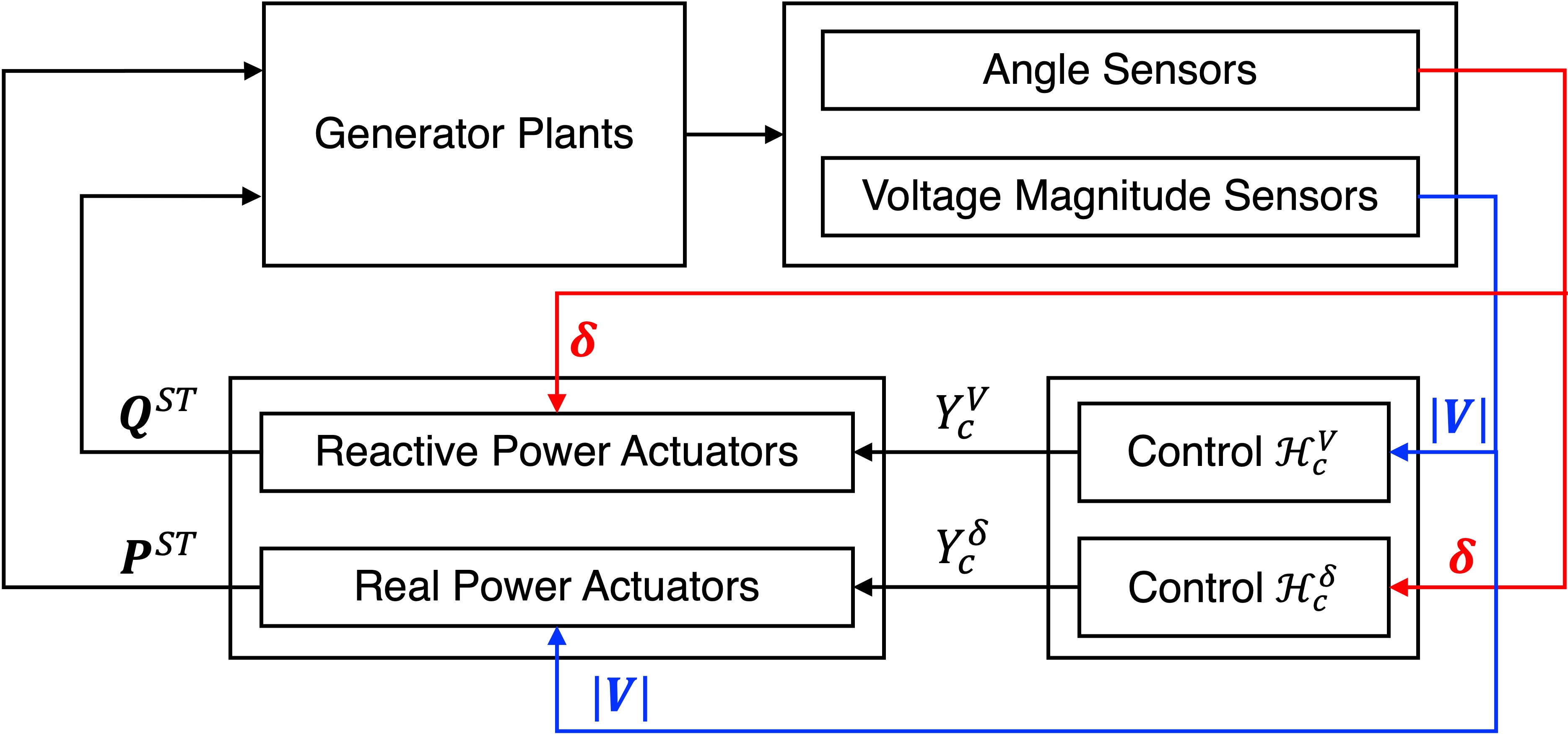}
		\caption{The control architecture.}
		\label{fig:control_architecture}
        \vspace{-0.5em}
	\end{figure}

	In what follows, we propose an angle and voltage magnitude feedback control framework as illustrated in Fig.~\ref{fig:overall_fb} and a control architecture as depicted in Fig.~\ref{fig:control_architecture}, where angle dynamics and voltage dynamics can be decoupled into two loops. The angle dynamics loop can be regarded as the positive feedback interconnection $(\widehat{\mathcal{H}}_{p}^{\delta}, \mathcal{H}_{c}^{\delta})^{+}$ of NI node plants $\mathcal{H}_{p}^{\delta}$ and OSNI edge controllers $\mathcal{H}_{c}^{\delta}$ based on the transmission network, while the voltage dynamics loop is considered as the negative feedback interconnection $(\widehat{\mathcal{H}}_{p}^{V}, \mathcal{H}_{c}^{V})^{-}$ of passive node plants $\mathcal{H}_{p}^{V}$ and output strictly passive edge controllers $\mathcal{H}_{c}^{V}$  based on the transmission network.
	
	\subsubsection{\bf Decoupled Angle Dynamics} 
	
	 In the angle dynamics loop (red dotted box in Fig.~\ref{fig:overall_fb}), we define $x_{pi}^{\delta} = [\dot{\widetilde{\delta}}_{i},  \widetilde{\delta}_{i} ]^{\top} \in \mathbb{R}^{2}$ and $u_{pi}^{\delta} \in \mathbb{R}$. The node plant of each synchronous generator bus $i \in \mathcal{V}$ is described by
	\begin{subequations}
		\label{eq:hpi_delta}
		\begin{align}
			H_{pi}^{\delta} \ : \quad	\dot{x}_{pi}^{\delta} &= A_{pi}^{\delta}x_{pi}^{\delta} + B_{pi}^{\delta}u_{pi}^{\delta}, \label{eq:hpi_delta_x}\\
			y_{pi}^{\delta} & = C_{pi}^{\delta}x_{pi}^{\delta}, \label{eq:hpi_delta_y} 
		\end{align}
	\end{subequations}
	where system matrices  are
	$
	A_{pi}^{\delta} = \begin{bmatrix}
		\frac{-D_{i}}{M_{i}} & 0 \\
		1 & 0 
	\end{bmatrix}, \ B_{pi}^{\delta} = \begin{bmatrix}
		\frac{1}{M_{i}} \\
		0
	\end{bmatrix}, \text{ and } \ C_{pi}^{\delta} = \begin{bmatrix}
		0 & 1
	\end{bmatrix}.
	$
	Each  edge controller $l \in \mathcal{L}$ for the transmission line $e_{l}$ is designed as 
	\begin{subequations}
		\label{eq:hcl_delta}
		\begin{align}
			H_{cl}^{\delta}: \quad	\dot{x}_{cl}^{\delta} &= 
			-\frac{1}{\tau^{\delta}_{l}}x_{cl}^{\delta} + \frac{K^{\delta}_{l[1]}}{\tau^{\delta}_{l}}u_{cl}^{\delta}, \\
			y_{cl}^{\delta} 
			&= x_{cl}^{\delta} - K^{\delta}_{l[2]}u_{cl}^{\delta}, 
		\end{align}
	\end{subequations}
	where $\tau^{\delta}_{l} >0$ and $K^{\delta}_{l[2]} > K^{\delta}_{l[1]} >0$. 
	
	Based on such a design, each   edge controller $l \in \mathcal{L}$ can  be equivalently implemented by putting two node controllers at the two end of the underlying transmission line $e_{l}$. Considering all transmission lines, the control action of the total real power output actuated by the large-scale battery at each synchronous generator bus $i \in \mathcal{V}$ is described by
	\begin{equation*}
			P_{i}^{ST}= \sum_{l \in \mathcal{L}}q_{il}y_{cl}^{\delta} - \sum_{j \in \mathcal{N}{i}}B_{ij}\left(V_{nom}^{2}\sin\overline{\delta}_{ij}
		- \vmi \vmj \sin \delta_{ij} \right),
	\end{equation*}
	where  $q_{il}$ is the $l$-th element in the $i$-th row of the incidence matrix $\mathbf{Q}$.
	\medskip
	
	\begin{theorem}\label{thm:angle_loop}
		Consider node plants $H_{pi}^{\delta}, i \in \mathcal{V}$ described by~\eqref{eq:hpi_delta} and edge controllers $H_{cl}^{\delta}, l \in \mathcal{L}$ described by \eqref{eq:hcl_delta}. Consider the positive feedback interconnection $(\widehat{\mathcal{H}}_{p}^{\delta}, \mathcal{H}_{c}^{\delta})^{+}$ of node plants and edge controllers based on the underlying transmission network. Then, the positive feedback system $(\widehat{\mathcal{H}}_{p}^{\delta}, \mathcal{H}_{c}^{\delta})^{+}$ achieves output consensus.
	\end{theorem}
	\medskip
    {\it Proof.}  We designate the storage functions of  each node system  and each edge controller as 
	$
	S_{pi}^{\delta} = \frac{M_{i}}{2}\dot{\widetilde{\delta}}_{i}^{2} $ and $S_{cl}^{\delta} = \frac{(x_{cl}^{\delta})^{2}}{2K^{\delta}_{l[1]}}$. For the positive feedback system $(\widehat{\mathcal{H}}_{p}^{\delta}, \mathcal{H}_{c}^{\delta})^{+}$, we can verify the criterion in Theorem~\ref{thm2:positive_feedback}, demonstrating its applicability. \hfill$\square$

	\begin{remark}
		Despite the fact that  edge controllers in  Theorem~\ref{thm:angle_loop} are presented in the linear form~\eqref{eq:hcl_delta},  as long as OSNI edge controllers are applied to the angle dynamics loop, the positive feedback system $(\widehat{\mathcal{H}}_{p}^{\delta}, \mathcal{H}_{c}^{\delta})^{+}$ achieves output consensus. Since $
		\widetilde{\delta}_{i} - \widetilde{\delta}_{j} \to 0$, we obtain that 	$\delta_{i} - \delta_{j} \to \overline{\delta}_{i} - \overline{\delta}_{j}$. Furthermore,  the work of \cite{chen2023design} proved the internal stability of the positive feedback system $(\widehat{\mathcal{H}}_{p}^{\delta}, \mathcal{H}_{c}^{\delta})^{+}$, i.e., $\dot{\tilde{\delta}}_{i}, i \in \mathcal{V} \to 0$.   Thus, our proposed OSNI edge controllers for the angle dynamics loop has the advantages of synchronizing bus frequencies and  retaining angle differences over transmission lines as steady-state values before faults.
	\end{remark}

	\subsubsection{\bf Decoupled voltage dynamics}
	In the voltage dynamics loop (blue dotted box in Fig.~\ref{fig:overall_fb}), we define $x_{pi}^{V} = \tvmi \in \mathbb{R}$ and $u_{pi}^{V} \in \mathbb{R}$.  The node plant of each synchronous generator bus $i \in \mathcal{V}$ is described by
	\begin{subequations}
		\label{eq:hpi_v}
		\begin{align}
			H_{pi}^{V} \ : \quad	\dot{x}_{pi}^{V} &= A_{pi}^{V}x_{pi}^{V} + B_{pi}^{V}u_{pi}^{V}, \label{eq:hpi_v_x}\\
			y_{pi}^{V} & = C_{pi}^{V}x_{pi}^{V}, \label{eq:hpi_v_y} 
		\end{align}
	\end{subequations}
	where 
	$
	A_{pi}^{V} = -\frac{\alpha_{i}}{\gamma_{i}}, \ B_{pi}^{V} = \frac{1}{\gamma_{i}}, \text{ and } \ C_{pi}^{V} = 1.
	$
	Each  edge controller $l \in \mathcal{L}$ for the transmission line $e_{l}$ is designed as 
	\begin{subequations}
		\label{eq:hcl_v}
		\begin{align}
			H_{cl}^{V}: \quad	\dot{x}_{cl}^{V} &= 
			-\frac{1}{\tau^{V}_{l}}x_{cl}^{V} + \frac{K^{V}_{l[1]}}{\tau^{V}_{l}}u_{cl}^{V}, \\
			y_{cl}^{V} 
			&= x_{cl}^{V} 
		\end{align}
	\end{subequations}
	where $\tau^{V}_{l} >0$ and $K^{V}_{l[1]} >0$.

	Based on such a design, each edge controller $l \in \mathcal{L}$ can be equivalently implemented by putting two node controllers at the two ends of the underlying transmission line $e_{l}$. Considering all transmission lines, the control action of the total reactive power output actuated by the large-scale battery at each synchronous generator bus $i \in \mathcal{V}$ is described by
	\begin{align}
		Q_{i}^{ST} = &{\vmi} \Big(  \sum_{j \in \mathcal{N}{i}}B_{ij}\left( V_{nom}\cos\overline{\delta}_{ij}
		- \vmj \cos\delta_{ij} \right)\notag \\
		&-\sum_{l \in \mathcal{L}}q_{il}y_{cl}^{V}  \Big) ,
	\end{align}
	where  $q_{il}$ is the $l$-th element in the $i$-th row of the incidence matrix  $\mathbf{Q}$.
	\medskip

	\begin{theorem}\label{thm:voltage_loop}
		Consider node plants $H_{pi}^{V}, i \in \mathcal{V}$ described by~\eqref{eq:hpi_v} and edge controllers $H_{cl}^{V}, l \in \mathcal{L}$ described by \eqref{eq:hcl_v}. Consider the negative feedback interconnection $(\widehat{\mathcal{H}}_{p}^{V}, \mathcal{H}_{c}^{V})^{-}$ of node plants and edge controllers based on the underlying transmission network. Then, the negative feedback system $(\widehat{\mathcal{H}}_{p}^{V}, \mathcal{H}_{c}^{V})^{-}$ achieves output consensus.
	\end{theorem}
	\medskip

    {\it Proof.}  We designate the storage functions of each node system and each edge controller as $S_{pi}^{V} = \frac{\gamma_{i}}{2}(x_{pi}^{V})^{2}$  and $S_{cl}^{V} = \frac{\tau^{V}_{l}}{2K^{V}_{l[1]}}(x_{cl}^{V})^{2}$. For the negative feedback system $(\widehat{\mathcal{H}}_{p}^{\delta}, \mathcal{H}_{c}^{\delta})^{-}$, we can verify the criterion in Theorem~\ref{thm:negative_feedback}, demonstrating its applicability. \hfill$\square$

	\begin{remark}
		Output consensus for the negative feedback system $(\widehat{\mathcal{H}}_{p}^{V}, \mathcal{H}_{c}^{V})^{-}$ means that $\overline{U}_{c}^{V} \equiv \overline{\widehat{Y}}_{p}^{V} = 0$. According to each edge controller \eqref{eq:hcl_v}, we obtain that $\overline{y}_{cl}^{V} =\overline{x}_{cl}^{V} = K^{V}_{l[1]}\overline{u}_{cl}^{V} = 0,$ which leads to that $\overline{U}_{p} \equiv (\mathbf{Q} \otimes I_{m})\overline{Y}_{c} = 0.$ Since $A_{pi}^{V} < 0$ and $\overline{u}_{pi} = 0$, each node plant~\eqref{eq:hpi_v} reaches zero steady state; i.e., $\overline{\tvmi} = \vmi - V_{nom} =  0 $.   Thus our proposed output strict edge controllers for the voltage dynamics has the advantage of regulating bus voltages.
	\end{remark}

	\medskip
	
	In summary, the proposed feedback control framework decouples the angle dynamics and voltage dynamics into two loops. As illustrated in Fig.~\ref{fig:control_architecture}, voltage phasor measurements are assumed, and large-scale batteries equipped at synchronous generator buses are utilized as control actuators, where real powers control the angle dynamics, and reactive powers control the voltage dynamics. 	 In particular, the proposed feedback control framework can guarantee frequency synchronization, angle differences preservation, and voltage regulation for electrical power systems.

	\section{Simulations}\label{sec:simulation}
\begin{figure}[bt]
		\centering
		\includegraphics[width=0.5\linewidth]{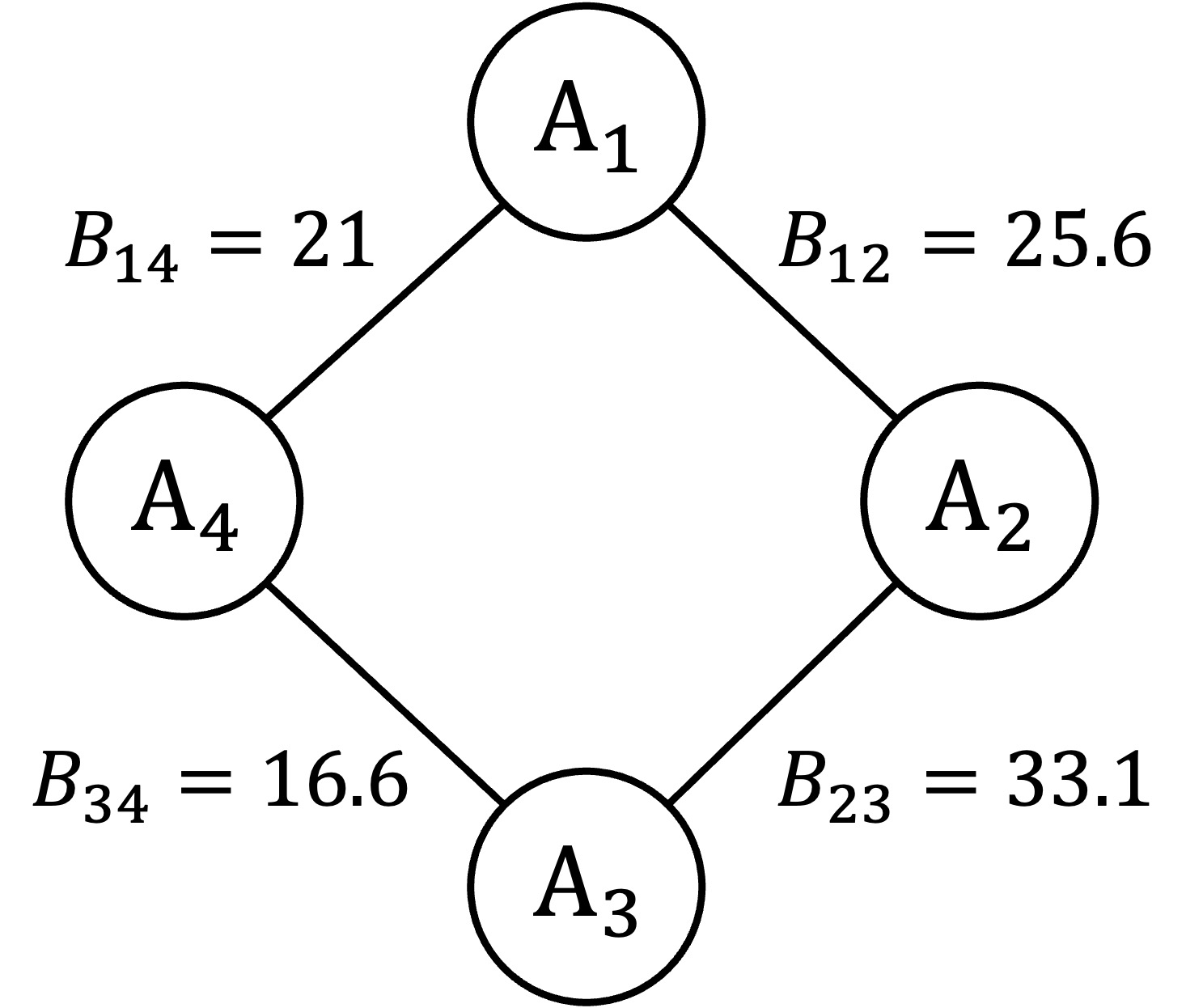}
		\caption{A four area equivalent network, where A$_{i}, i \in \mathcal{V}$ represents area $i$, and $B_{ij}, (i,j) \in \mathcal{E} $ represents the susceptance of the transmission line $(i,j)$ \cite{trip2016internal, nabavi2013topology}.}
		\label{fig:4-generator-bus}
        \vspace{-0.5em}
	\end{figure}

	Consider a connected four-area-equivalent transmission network as illustrated in Fig.~\ref{fig:4-generator-bus}, which is obtained for the South Eastern Australian 59-bus system \cite{nabavi2013topology}. The areas A$_1$ and A$_2$ act as generators, while the areas A$_3$ and A$_4$ act as motor loads. The system parameters and controller settings used in the simulations are summarized in Table~\ref{tab:numerical_values} and Table~\ref{tab:controller_parameters}, respectively.   \medskip
	
	\begin{table}[bt]
		\centering
		\begin{tabular}{lllll}
			\hline
			& Area 1 & Area 2 & Area 3 & Area 4\\
			\hline
			\multicolumn{5}{c}{System parameters}\\
			\hline
			$M_{i}$   &  5.22& 3.98& 4.49& 4.22\\
			$D_{i}$   &  1.6& 1.22& 1.38& 1.42\\
			$T'_{doi}$  &  5.54& 7.41& 6.11& 6.22\\
			$X_{di}$   &  1.84& 1.62& 1.8& 1.94\\
			$X'_{di}$   &  0.25& 0.17& 0.36& 0.44\\
			$B_{ii}$   &    -49.61 & -61.66 & -52.17&-40.18\\
			\hline
			\multicolumn{5}{c}{Equilibrium}\\
			\hline
			$\overline{P}^{G}_{i}$ &    8.076 &12.04&-14.38 & -5.735\\
			$\overline{E}^{ex}_{i}$ &      7.824 & 9.13 & 8.437 & 6.864 \\
			$\overline{\delta}_{i}$ &  30& 28& 5& 10 \\
			$\overline{\omega}_{i}$ &  50& 50& 50& 50 \\
			$\overline{|V_{i}|}$ &  1 & 1 & 1& 1\\ 
			\hline
			\multicolumn{5}{c}{Initial Deviation }\\
			\hline
			$\widetilde{\delta}_{i}$ &  10& -8& -3& -10 \\   
			$\dot{\widetilde{\delta}}_{i}$ & 0 & 0 & 0 & 0  \\
			$\widetilde{|V_{i}|}$ &  0.04 &-0.04 & -0.05 & 0.05  \\
			\hline
		\end{tabular}
		\caption{System parameters are provided in per unit.}
		\label{tab:numerical_values}
         \vspace{-0.5em}
	\end{table}
	
	\begin{table}[bt]
		\centering
		\begin{tabular}{lllll}
			\hline
			& Line 12 & Line 23 & Line 34 & Line 41\\
			\hline
			\multicolumn{5}{c}{Angle controllers}\\
			\hline
			$\tau_{l}^{\delta}$   &  1& 1& 1& 1\\
			$K_{l[1]}^{\delta}$   &  0.4& 0.5& 0.3& 0.4\\
			$K_{l[2]}^{\delta}$  &  0.7& 0.8& 0.6& 0.8\\
			\hline
			\multicolumn{5}{c}{Voltage magnitude controllers}\\
			\hline
			$\tau_{l}^{V}$   &  1& 1& 1& 1\\
			$K_{l[1]}^{V}$   &  0.4 & 0.5 & 0.3 & 0.4\\
			\hline
		\end{tabular}
		\caption{Controller parameters.}
		\label{tab:controller_parameters}
     \vspace{-0.5em}
	\end{table}
	
\begin{figure}[bt]
		\centering
		\includegraphics[width=0.85\linewidth]{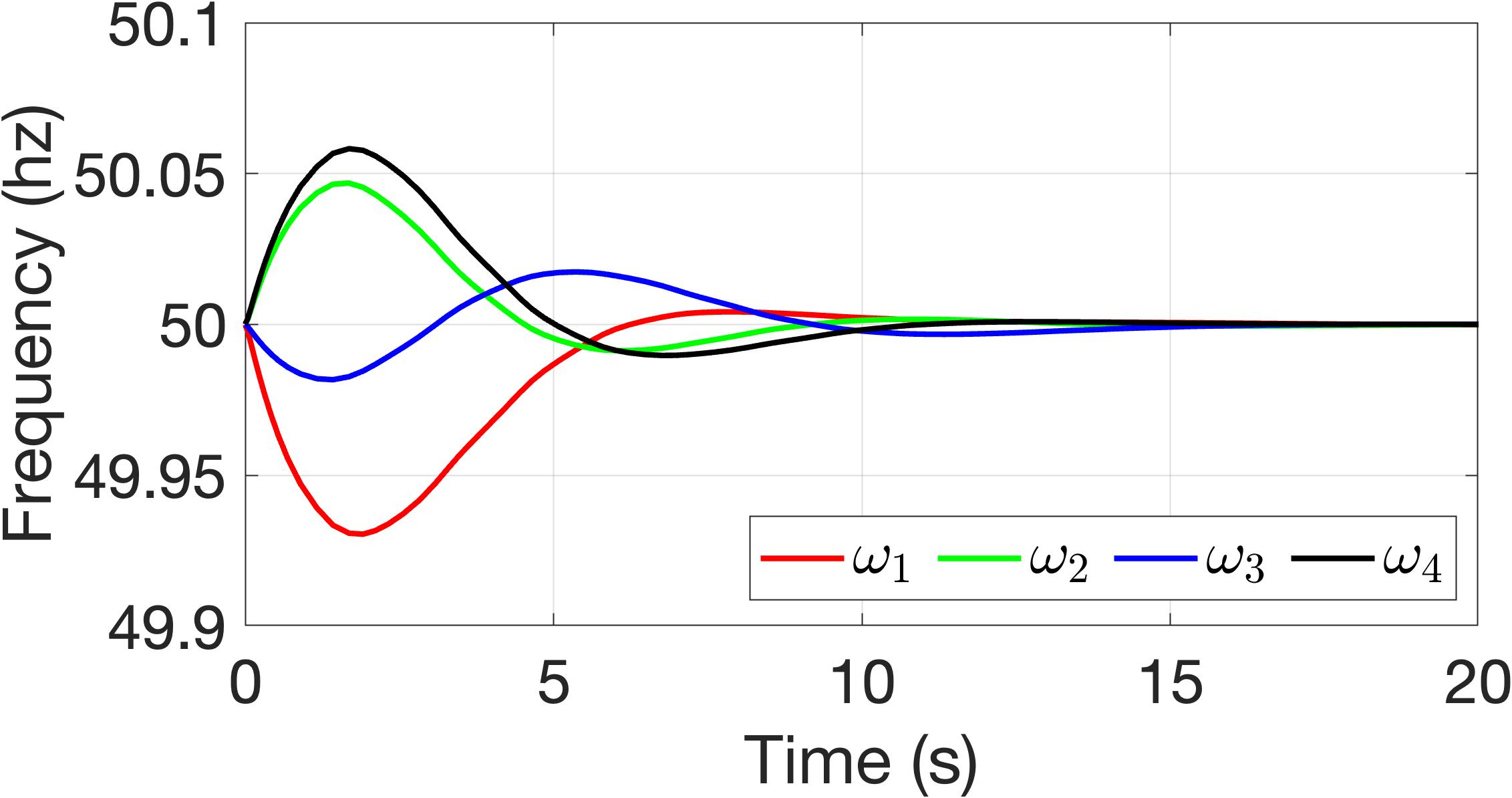}
		\caption{Frequencies of generator buses.}
		\label{fig:frequency_synchronization}
	\end{figure}

\begin{figure}[bt]
		\centering
		\includegraphics[width=0.85\linewidth]{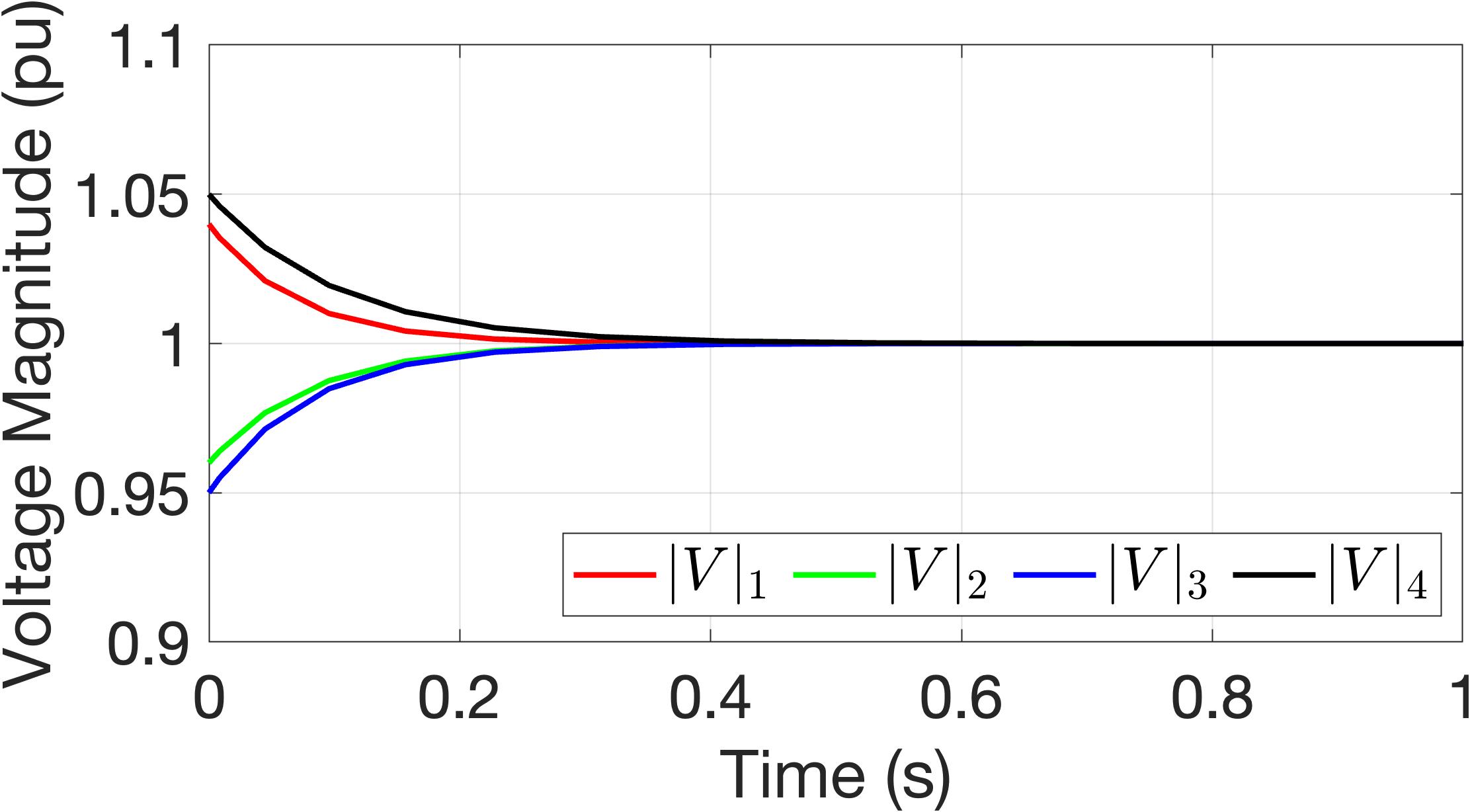}
		\caption{Voltage magnitudes of generator buses.}
		\label{fig:voltage_regulation}
	\end{figure}

\begin{figure}[bt]
		\centering
		\includegraphics[width=0.85\linewidth]{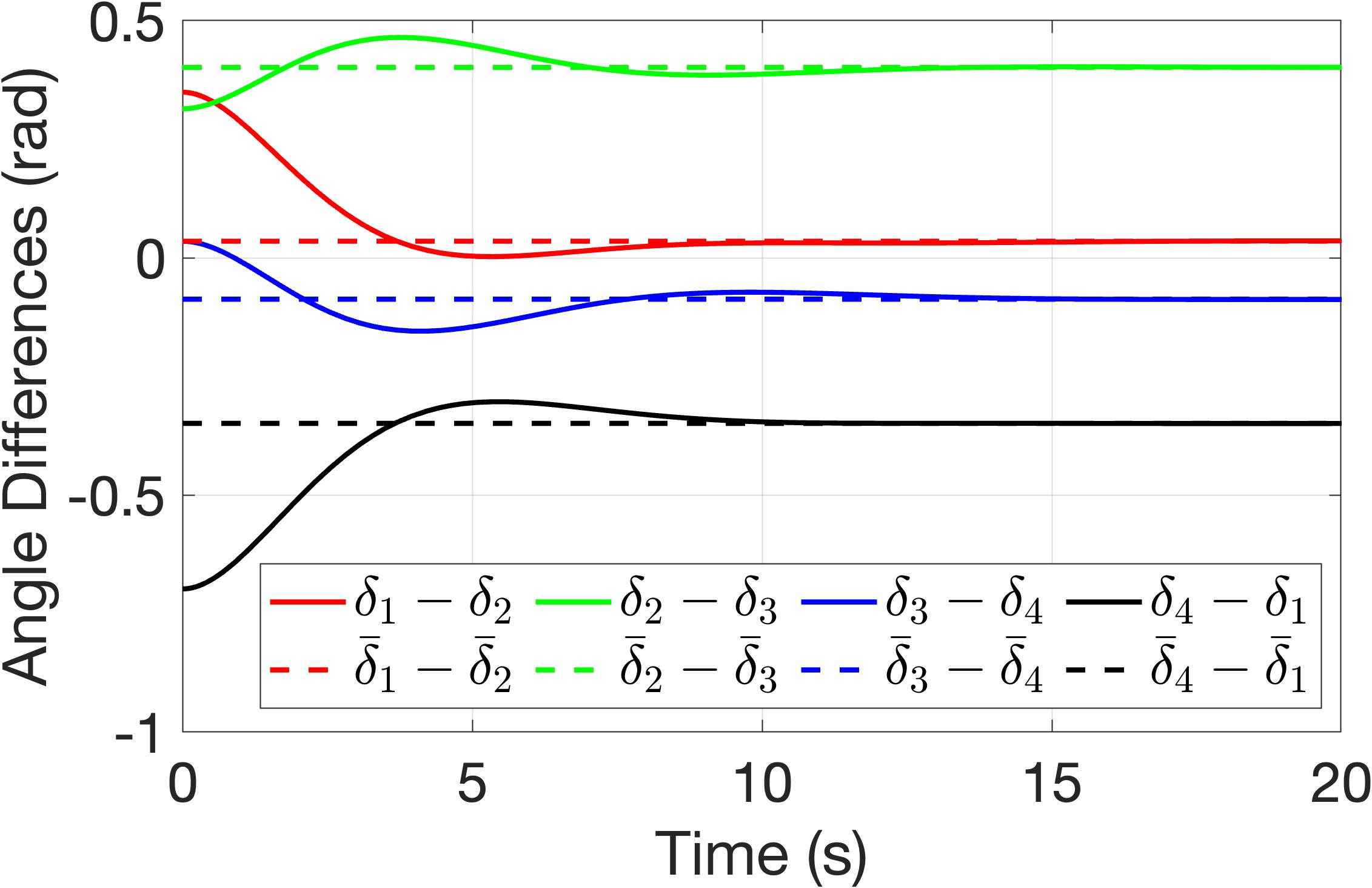}
		\caption{Angle differences over transmission lines.}
		\label{fig:angle_difference_preservation}
      \vspace{-0.5em}
	\end{figure}

	{\bf \noindent Simulation Results.} First, we plot generator bus frequencies in Fig.~\ref{fig:frequency_synchronization}. Under our proposed controllers, the bus frequencies are synchronized at the nominal value of $50$ Hz. Second, we present the comparison of $\delta_{i}-\delta_{j}$ and $\overline{\delta}_{i}-\overline{\delta}_{j}$ for all $(i,j) \in \mathcal{E}$ in Fig.~\ref{fig:angle_difference_preservation}.  It is shown that using our designed controllers, the angle differences $\delta_{i}-\delta_{j}$ for all $(i,j) \in \mathcal{E}$ are maintained at the corresponding steady-state angle differences at the equilibrium $\overline{\delta}_{i}-\overline{\delta}_{j}$ for all $(i,j) \in \mathcal{E}$.  Third, we plot voltage magnitudes of generator buses in Fig.~\ref{fig:voltage_regulation}. With our controllers, the bus voltage magnitudes are regulated at the desired value $\bvmi = 1, \forall i \in \mathcal{V}$. In summary, the aforementioned results validate Theorems~\ref{thm:angle_loop}-\ref{thm:voltage_loop} and demonstrate three advantages of our proposed control controllers: frequency synchronization, angle difference preservation, and voltage regulation.

	
	\section{Conclusion}\label{sec:conclusion}
    This paper presented a unified approach to tackle voltage regulation, frequency synchronization, and rotor angle stabilization in power grids. We formulated our problem as an output consensus problem and proposed a passivity and negative-imaginary based control framework. By leveraging real-time voltage phasor measurements, we showed that large-scale batteries co-located at synchronous generators could serve as actuators for the control of real and reactive power for voltage, frequency, and rotor angle regularization. Simulation results confirmed the efficiency and robustness of our proposed approach. Incorporating the saturation of battery actuators is a possible future research direction.

	\bibliographystyle{IEEEtran}
	\bibliography{ref}
	
\end{document}